\documentclass[twoside]{article}

\usepackage{PRIMEarxiv}

\usepackage[utf8]{inputenc} % allow utf-8 input
\usepackage[T1]{fontenc}    % use 8-bit T1 fonts
\usepackage{hyperref}       % hyperlinks
\usepackage{url}            % simple URL typesetting
\usepackage{booktabs}       % professional-quality tables
\usepackage{amsfonts}       % blackboard math symbols
\usepackage{nicefrac}       % compact symbols for 1/2, etc.
\usepackage{microtype}      % microtypography
\usepackage{subcaption}
\usepackage{lipsum}
\usepackage{fancyhdr}       % header
\usepackage{graphicx}       % graphics
\usepackage{siunitx}
\graphicspath{{media/}}     % organize your images and other figures under media/ folder
\usepackage{cancel}
\usepackage{multicol}
\usepackage{hyperref}
\usepackage{booktabs}
\usepackage[utf8]{inputenc}
\usepackage[cmex10]{amsmath}
\usepackage{algorithm}
\usepackage{algorithmic}
\renewcommand{\figurename}{Fig.}
\renewcommand{\tablename}{Table}

\newcommand{\figref}[1]{\textcolor{blue}{\figurename~\ref{#1}}}
\newcommand{\tabref}[1]{\textcolor{blue}{\tablename~\ref{#1}}}

\hypersetup{
    colorlinks=true,
    linkcolor=blue,
    citecolor=blue,
    filecolor=blue,
    urlcolor=blue
}

\usepackage{multirow}

\title{Systematic Gray-Box Identification Methodology for Voltage Source Converters
\thanks{\textit{\underline{\textbf{Submitted} for publication in IEEE Transactions on Power Delivery}} \\} 
}

\author{Nicolae Darii, Luis A. Garcia-Reyes, Ignasi Ventura Nadal,\\
Oscar Sabor\'{i}o-Romano, Ranjan Sharma,\\
Oriol Gomis-Bellmunt, and Nicolaos A. Cutululis}

\begin{document}
\maketitle

\begin{abstract}
This paper introduces a systematic gray-box identification framework for voltage-source converter models based solely on terminal time-series data. The proposed approach combines a physically informed white-box standard model with iterative time-domain calibration to estimate controller parameters that mimic the behavior of the black-box model in electromagnetic transient (EMT) simulations. Unlike conventional frequency-domain identification methods, the framework leverages time-domain data more effectively to better constrain the surrogate model across a broader operating range and capture reference-signal dynamics. To evaluate the accuracy of the identified model, the paper presents additional frequency-domain validation metrics based on Nyquist analysis and singular value decomposition, allowing for both quantitative assessment of model divergence and qualitative classification of mismatch types. The methodology is tested on cases with increasing structural uncertainty, from exact parametric recovery to an actual detailed  EMT  black-box model. Results demonstrate that the proposed framework can accurately recover parameters when the internal structure is known, adjust for moderate structural mismatch with extra degrees of freedom, and offer a reliability measure for small-signal stability analysis of converter models protected by intellectual property.
\end{abstract}

% keywords can be removed

\section{Introduction}
The rise of renewable energy and Inverter-Based Resources (IBRs) is transforming traditional power systems, with IBRs worldwide expected to contribute 43\% of electricity generation by 2030 \cite{RenewableIEA}. Large IBR plants, like the 3.6 GW Dogger Bank Offshore Wind Power Plant, will require multi-vendor collaboration \cite{dogg}. As IBRs can induce dynamic instability, it is essential to employ system analysis tools such as  Small-Signal Analysis (SSA) or transient stability analysis, to ensure stability and proper operation in modern power systems \cite{Mugambi2025MethodologiesAnalysis}.

% Focus on the specific issue
Thus, the need to adopt stability analysis via SSA or through impedance-based methods when detailed modes are not available (all practical industrial cases), alongside Electromagnetic Transient (EMT) simulations, is becoming increasingly critical. This can ensure the so-called interoperability among power system users \cite{Efficiency2025FosteringSystem}, because frequency-domain (FD) methods can aid in a more comprehensive understanding of the phenomena and effects involved. However, preserving the Intellectual Property (IP) of Original Equipment Manufacturers (OEMs) remains essential to maintain competitiveness among IBR players \cite{Ivanov2016PrescriptionSolutions}. This dichotomy creates a fundamental conflict between the necessity of knowing as much as possible about the IBR's structure to model and conduct studies in FD, and the need not to disclose as much as possible to protect the IP. Therefore, this paper's research question addresses how to derive physically meaningful surrogate models from IP-protected converters' models for stability studies when FD is not provided alongside the black-box (BB) EMT model. The main reason is the possibility of better interpreting the model and the phenomena when the model is open, as the surrogate one is.

% What is needed and the definition of terms
\subsection{Gray-Box Concept}
To address such a problem, it could be helpful to practice extracting information "at the terminals" of the internally unknown IBR models, also known as black-box (BB) models. Therefore, formulate an educated guess (ansatz) of the potential internal control structure of the BB, and build a white-box (WB) model. Finally, select a set of WB parameters that will serve as degrees of freedom to make the WB model behave like the BB at the terminals. This surrogate model can then be used for more in-depth studies, and it is defined as a gray-box (GB) model.

% Possible mathematical tools & similarities
The extraction practice is rooted in system identification and GB modeling theory, a field well defined in the literature \cite{Sohlberg1998GreyModelling}. In power system applications, surrogate models have already been adopted for many electrical machines (e.g., transformers, synchronous generators). A base structure is defined, and its electrical parameters are then determined through several tests on the actual machine. Although not the exact model of the actual machine, a surrogate standard model can be tuned to match the specific purpose. The extension to IBRs is feasible, where, differently, it is necessary to create a surrogate base structure for the control system, principally because it is the core of its principal behavior.

% Literature context
\subsection{Related work}

In the power system, various GB applications exist. For example, in \cite{Zhang2024DominantConverters}, it is indicated that, without prior knowledge of the internal structure, it is possible to identify dominant modes using frequency responses extracted from frequency scans. Particle swarm optimization (PSO), a metaheuristic method, performs vector fitting (VF) on a test transfer function, applying the ansatz to the order. A similar optimization method is used in \cite{Pan2025GreySensitivity, Vilera2024ControlOptimization}. In \cite{Zong2021GreyAnalysis}, the overall Doubly Fed Induction Generator-based Wind Turbine (WT) aggregated plant is modeled with an analytical transfer function, using designed nominal parameters as degrees of freedom (8 electrical, 12 control). In \cite{Amin2019AImpedance}, a GB technique fits analytical FD VSC structures to BB frequency responses, limiting parameter estimation to determine affected bandwidths. The reference frequency response is obtained from EMT model frequency scans, with a two-step, iterative, nonlinear optimization tool used for VF. Ref. \cite{Zhu2022ParticipationStability} extracts GBs at the network level, establishing a direct link between the system's eigenvalues and impedance sensitivity to system parameters. The reference information from the BB system is the frequency-dependent impedance scanned at target nodes. Ref. \cite{Cecati2025InteroperabilityPerspective} introduces partially nonlinear blocks into the WB model, implemented as transfer functions to capture unknown extra dynamics. This addition produces a modified transfer function that performs VF to extract information from BB systems, emphasizing its importance in enhancing vendor interoperability and leading to stricter tuning guidelines for suppliers by obtaining additional information about BB converter models. In \cite{Li2023ADiagrams}, the converter's frequency response is synthesized into a discrete numerical transfer function, later used for the Nyquist Stability criterion, also applied in \cite{Haugaard2024Immittance-basedComponents}. Here, the GB is the numerical transfer function at a precise operational condition. Similarly, ref. \cite{Garcia-Reyes2025Data-DrivenConverters} synthesizes an equivalent state-space model with model-order reduction and improved pole localization, while \cite{Garcia-Reyes2026EnablingSystems} introduces the adaptive pole-expansion mechanism and its application to large-scale SSA, enabling black-box‑dominated stability assessment.
 In \cite{Helman2024Grey-boxInverters}, the GB approach is applied to the grid-forming (GFM) converter structure by fitting it to a nonlinear swing equation and identifying parameters to determine equivalent inertia, damping, and droop that match the GMF FD response.

\subsection{Research Gap and Proposed Approach}
The existing FD-based method guarantees terminal equivalence at the linearization point and for the specific excitation signals used during identification (usually just voltage or current), leading to an equivalent FD impedance. However, power system studies require models that generalize across operating conditions and respond correctly to reference signals, which are rarely used in FD-based extraction. Time-domain identification effectively addresses this issue by using a single dataset with multiple excited reference signals to constrain the GB model across a wider range of operating conditions. Additionally, time-domain data is much easier to obtain in industrial environments. For example, a step response can be captured without specialized frequency scanning, either in reality or via software.

\begin{figure}[t!]
    \centering
    \includegraphics[width=0.9\linewidth]{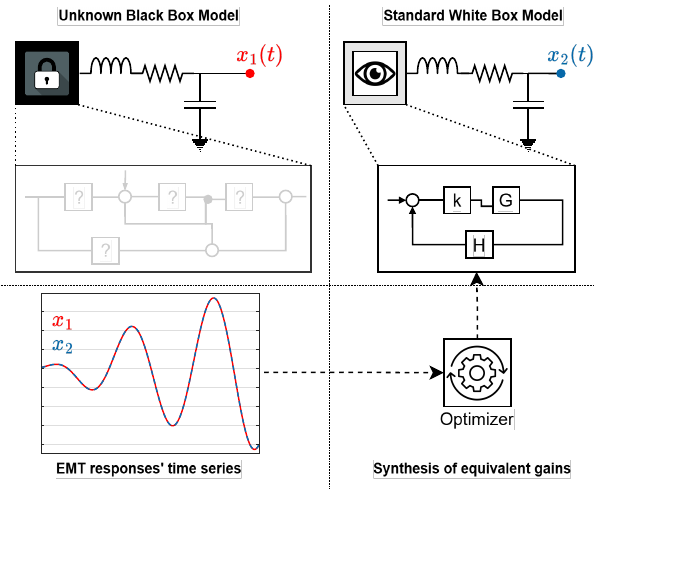}
    \caption{Gray-Box concept.}
    \label{fig: concept}
\end{figure}

The work proposes a systematic EMT-based GB identification framework for IBRs that, after selecting a potentially suitable model structure backbone, relies solely on terminal measurements conceptualized in 
\figref{fig: concept} to fit the chosen open model to behave like the BB one at the terminals. Additionally, we introduce a dual quantification and qualification methodology to evaluate the fidelity of the identified GB model, thereby providing a robust metric for the reliability of the resulting SSA.

The contributions of this paper are as follows:

\begin{itemize}
    \item Unified GB identification workflow integrating EMT simulations and FD validation
    \item A novel FD domain metric based on Singular Values Decomposition (SVD) to quantify model reliability
    \item A systematic analysis that applies the GB method to models with a gradual level of internal structure uncertainty. From known structure and unknown internal gains to a detailed but black-boxed EMT model.
    \item An open-source gray-box framework routine publicly available in~\cite{Darii2026Grey-boxAlgorithm}.
\end{itemize}

\section{Methodology}
\label{Methods}
\begin{figure}[t!]
    \centering
    \includegraphics[width=0.9\linewidth]{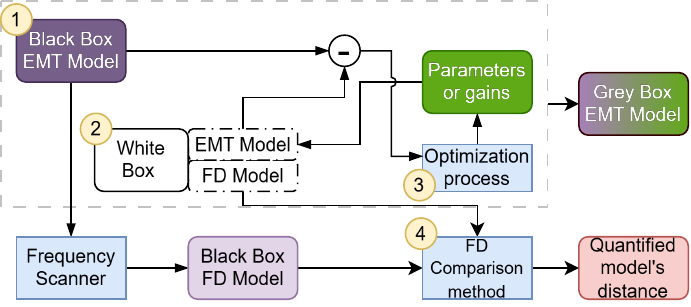}
    \caption{Proposed methodology for developing gray-box equivalent models.}
    \label{fig:GBscheme}
\end{figure}

This section provides a detailed description of the gray-boxing process. The extraction is defined by two workflows: 1) fitting a fully known WB model to the BB's time-domain behavior and 2) evaluating the distance between the BB and the GB in FD, as shown in \figref{fig:GBscheme}.

The two workflows are then combined by sub-processes unfolded as:

\begin{enumerate}
    \item Black-box setup and reference generation (\ref{F1})
    \item Educated guess (ansatz) of the gray-box structure (\ref{F2})
    \item Iterative gray-box model calibration (\ref{F3})
    \item  Model-to-data fidelity assessment (\ref{F4})
\end{enumerate}

%2-  the writer provides background information and justification: background delle varie definizioni elementi e tools usati
The required tools are: EMT software for manipulating time-domain VSC models, Frequency Scanners for extracting numerical frequency responses from the models, and an optimization algorithm that can be interfaced with the EMT simulation. In principle, the time-series and frequency-response of the BB do not strictly require ownership of the model; they can work with data previously extracted and readily shared in tabular formats. However, the emerging trend of sharing EMT-encrypted files in Dynamic Link Library (DLL) format warrants a description of the procedure for direct BB information extraction.

However, it must be emphasized that any such work or extraction must be carried out within the legal framework governing the BB models, while respecting all applicable non-disclosure agreements. As this paper focuses on technical aspects, these requirements are acknowledged, but not discussed in detail. 

%3- The writer provides an overview of the procedure/method itself.: passo passo dettagliato come si fa il greyboxing

\subsection{Black-Box Setup and Reference}
\label{F1}
\begin{figure}[t!]
    \centering
    \hspace{7mm}
    \includegraphics[width=0.9\linewidth]{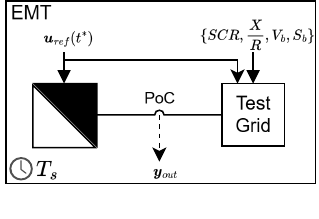}
    \vspace{-8mm}
    \caption{Black-box setup.}
    \label{fig: simulation_setup}
\end{figure}

The first step involves defining the output reference variables ($\mathbf{y}_\text{out}$), the excitation routine $\mathbf{u}_\text{ref}(t^*)$ for the input (duration and perturbation magnitude), and EMT's simulation time step ($\Delta t$). Then, the converter model must be connected to a test Thévenin Equivalent (Test Grid) with known parameters (short-circuit ratio SCR, XR-ratio, Nominal Voltage $V_\text{b}$, and Power $S_\text{b}$).
The choice of reference variables is open; in general, with BB models, only the output's terminal variables are accessible. The same exact Test Grid specifications, I/O variables, and simulation time step, $\Delta t$, must be noted and transposed in the WB's EMT model (later becoming GB) to replicate the exact benchmark shown in \figref{fig: simulation_setup}. 

The $\mathbf{u}_{\mathrm{ref}}(t) = \mathbf{A} \odot \mathbf{H}(t)$, where $\mathbf{A} = (A_i)_{i=1}^n$ is the vector of step magnitudes, and $\mathbf{H}(t) = \big(H(t - t_{0,i})\big)_{i=1}^n$ is the vector of Heaviside functions, so that each component of $\mathbf{u}_{\mathrm{ref}}(t)$ undergoes a step at its corresponding activation time $t_{0,i}$.

These must induce an output response $\mathbf{y}_\text{out}(t)$ rich enough to adequately excite the system modes and render them observable in the output dynamics. Increasing the amplitude or sharpness of the step excitation generally makes the modal contributions more explicit in the output response. The excitation must be large enough for the relevant modes to be observable, but not so large that it triggers protection functions or induces control‑mode transitions (e.g., from steady‑state voltage/reactive‑power control to fault‑ride‑through). Moreover, if the BB model includes rate limiters, these may distort the response. In such cases, the excitation should be applied from the Test Grid side, e.g., by perturbing the network voltage amplitude or phase angle, as illustrated in \figref{fig: simulation_setup}.

The same Test Grid setup is also used for frequency response extraction. The only difference is that there is no need to excite the system through $\textbf{u}_\text{ref}(t)$; instead, the same initial operating condition is selected for both the GB and BB models.
This choice is consistent with the requirement that the inputs applied to the BB model must be identically replicated in the GB model within the same experimental setup, ensuring a fair and meaningful comparison.

\subsection{Gray-Box Structure}
\label{F2}
Once the output time series $\mathbf{y}_\text{out}(t)$ from the BB EMT's model is determined, the GB structure must be defined. This step requires an educated guess (ansatz) that depends principally on the type of device (e.g., a wind turbine, STATCOM, PV, etc.), which generally has a standard structure \cite{Gaertner2020DefinitionTurbine}. This step significantly influences GB's success, since the more prior knowledge there is about the potential BB structure, the better the process will be. 

The selection of the GB's structure is followed by defining the GB's Degrees of Freedom (DoF), which are then used as a decision variable during the optimization process. In principle, every parameter of the IBR can be used as decision variables, as is done in \cite{Zong2021GreyAnalysis}. If certain variables are known in advance, the likelihood of successful GB is correspondingly increased. For example, in the case of IBRs, parameters of passive components such as filters, DC capacitors, inductors, and related elements may, in some instances, be obtained from OEM models or accompanying documentation. Therefore, the research recommends selecting controller gains $\mathbf{k} = \{k_1, k_2, \dots, k_i\}$ as DoF to facilitate the optimization process by reducing the number of decision variables and avoiding the inclusion of potentially unnatural passive parameters as optimization outputs.

\subsubsection{Small-signal analysis twin of the gray-box model}
The privilege of fully understanding the GB structure is an advantage for obtaining an exact, parametric, and validated Small Signal Model (SSM). The twin SSM serves as a support for performing SSA, including participation factor analysis for known states and parametric sensitivity analysis \cite{Fan2020Admittance-BasedAnalysis}. With additional physical information on the various model parameters, it provides a more complete picture than numerical SSMs, which are extracted directly through rational approximations. 
However, to build an SSM twin effectively, it is necessary to have an easily convertible GB model; therefore, the research avoided nonlinear elements that are not readily convertible, at least through the linearization of nonlinear differential equations, such as rate limiters.

\subsection{Iterative Time Domain Gray-Box Model Calibration}
\label{F3}

Once the GB structure is defined, the overall models should be integrated into an optimization routine. The choice of optimization tool can range from gradient descent (e.g., Sequential Quadratic Programming, interior-point methods) to metaheuristics (e.g., genetic algorithms, particle swarm optimization). The interior-point method was chosen for its robustness and efficiency, and was implemented using the MATLAB function \texttt{fmincon}  \cite{FminconMATLAB} as an example. However, since the procedure aims to provide a general framework, the optimizer could be replaced with one that suits the end-user case. 
By expressing the optimization process in a general analytical form, the GB-EMT simulation process can be defined as
\begin{equation}
\mathbf{y}_\text{out}(t) = f(\mathbf{k}, \mathbf{u}_\text{ref}(t), t) \, ,
\label{eqn:GB_general}
\end{equation}
with decision variables $\mathbf{k} = \{k_1, k_2, \dots, k_i\}$, and selected sets of reference inputs with clearly defined excitation times 
$\mathbf{u}_\text{ref} \in \mathbb{R}^{\text{n} \times \text{p}}$, where $\text{n}$ denotes the number of time steps and $\text{p}$ the number of input signals. The selected outputs at the terminals 
$\mathbf{y}_\text{out} \in \mathbb{R}^{\text{n} \times \text{m}}$, where $\text{m}$ is the number of measured output signals, are used in the objective function, where the cost function is:
\begin{equation}
J = 
\begin{cases} 
\sum_{i=1}^{n} \sum_{j=1}^{m} \left( y_{\text{out},i,j}^\text{BB} - y_{\text{out},i,j}^\text{GB} \right)^2 & \gamma=0 \\
\infty & \gamma=1
\end{cases}
\label{eqn:cost_general}
\end{equation}
being $s$ the converge criteria of $f$ and $\mathbf{y}_{\text{out}}^\text{BB} \, , \mathbf{y}_\text{out}^\text{GB} \in \mathbb{R}^{n \times m}$ the data series outputs from the BB and GB, respectively. Eq.~\eqref{eqn:cost_general} is used in the optimization process as
\begin{equation}
\mathbf{\hat{k}} = \arg\min_{\mathbf{k}} \, J \bigl( f(\mathbf{k}, \mathbf{u}_\text{ref}(t), t) \bigr) \, ,
\label{eqn:opt_general}
\end{equation}
which outputs the set of gains $\mathbf{\hat{k}}$ that, once applied in GB-EMT, emulate the terminal outputs of the BB.

The optimization problem becomes: since the cost function \eqref{eqn:cost_general} is derived from EMT simulations, the optimization process may test sets of gains that prevent the simulations from converging. For this reason, $J$ is a stepwise function that tends to infinity when the simulation $f$ does not converge $\gamma=1$) and equals the sum of squared errors when the simulation converges ($\gamma =0$). It is worth noting that, for the $\boldsymbol{k}$, the upper and lower bounds should be defined heuristically, considering typical tuning, then scaled along with the gains before optimization to roughly match in magnitude, and scaled back when applied to EMT. Additional potential constraints could be placed on the Bandwidth (BW) of the control loops so the gains exploration would look just for values that respect the inner and outer loop speed hierarchy, e.g. $\text{BW}_\text{in} > 10\text{BW}_\text{out} $ \cite{Amin2019AImpedance}).
Finally, the gain values extracted, $\mathbf{\hat{k}}$, can be used in the SSM-GB twin to perform a deeper stability analysis. Additionally, the initial point of the SSM can be directly obtained from the EMT-GB simulations. 

\section{Gray-Box Error Analysis}
\label{F4}
If the GB ansatz and the BB have the same structure, the process becomes the mere identification of the BB parameters. This is not the case when the GB and BB are structurally different. The fitting process may produce perfectly overlapping output responses with physically meaningful parameters, but differences may be observed across various frequency ranges when analyzed in the FD. Conversely, as the literature suggests, since FD fitting would yield a perfectly matching FD response but would lead to nonphysical gain values. 

For this reason, the research proposes using FD to quantify the distance between the BB and the extracted GB, and to use it as a quality metric for interpreting the SSA results on surrogate models. Additionally, it aims to identify whether the differences are due to natural frequencies, damping mismatch, or both.

\subsection{Nyquist Stability Criterion}
The first comparison method uses the widely known Nyquist plots to analyze the eigenvalues of the open-loop response between the converter and the Test Grid defined by:
\begin{equation}
    \label{eqn: nyquist}
    \boldsymbol{\lambda_i}({\boldsymbol{Y}_\text{c}(j\omega_i)\boldsymbol{Y}_\text{g}^{-1}(j\omega_i))} \quad i=1,...,n \, ,
\end{equation}

With $\boldsymbol{Y}_\text{c}$ being the converter $dq$-admittance tensor ($2\times 2\times n$ for a frequency range $n$) and $\boldsymbol{Y}_\text{g}$ of the Test Grid admittance tensor extracted through EMT impedance scans \cite{Garcia-Reyes2026SIaD-Tool:Systems}.

This can provide a quick visual approach to estimate whether the stability margins and general frequency behavior are coherent. Additionally, the frequency of the most undamped mode can be determined. This evaluation depends entirely on scanner density and may be inaccurate for more precise evaluations.

\subsection{Singular Value Decomposition}

The second method is widely used in robust control theory and explicitly defines the Multiple Input Multiple Output (MIMO) closed-loop frequency response
\begin{equation}
\label{eqn: closedloop}
    \boldsymbol{T}(j\omega_i) = (\boldsymbol{I} + \boldsymbol{Y}_\text{c}(j\omega_i)\boldsymbol{Y}_\text{g}^{-1}(j\omega_i))^{-1}(\boldsymbol{Y}_\text{c}(j\omega_i)\boldsymbol{Y}_\text{g}^{-1}(j\omega_i)) \, .
\end{equation}
Then it is computed the Singular Value Decomposition (SVD) for each frequency $i$ as 
\begin{equation}
\label{eqn: svd}
    \boldsymbol{T}(j\omega_i) = \boldsymbol{U}(j\omega_i)\,
    \begin{bmatrix}
        \sigma_1(j\omega_i) & 0 \\
        0 & \sigma_2(j\omega_i)
    \end{bmatrix}
    \boldsymbol{V}^\text{H}(j\omega_i) \, ,
\end{equation}
where $\mathbf{U}(j\omega_i)$ and $\mathbf{V}(j\omega_i)$ contain the left and right singular vectors, respectively. The singular values are real, non‑negative, and ordered as $\sigma_1(j\omega_i) \ge \sigma_2(j\omega_i) \ge 0$. The largest singular value, $\sigma_1(j\omega_i)$, corresponds to $\sigma_{\max}(j\omega_i)$ and captures the dominant information regarding the system performance. The main reason for moving to this domain is the need to switch from the rotating reference frame ($dq$) to a more convenient one, as it is hard to compare the two systems from the eight frequency response plots. 

The study of the maximum SVD value  $\sigma_{\max}(j\omega_i)$ at each frequency $i$ of the closed-loop transfer function is equivalent to representing the maximum system gain through the spectral norm of the system transfer function $G(j\omega)$:
\begin{equation}
\label{eqn: Hinfty}
    ||G(j\omega)||_{\infty} =\sup_{\omega\in\mathbb{R}}\sigma_{\max}(G(j\omega )) \, .
\end{equation}
Tensor reshaping in the real tensor $\mathbb{C}^{2\times2\times n}\rightarrow\mathbb{R}^{n}$ is possible due to the SVD property of synthesizing meaningful matrix information into the maximum singular value $\sigma_{\max}$. Additionally, it simplifies comparisons between GB and BB FD models. The SVD can also help locate the frequency of the dominant and less-damped modes through the Complex Mode Indication function (CMIF) defined as \cite{Shih1988ComplexEstimation}:
\begin{equation}
\label{eqn: CMIF}
    \text{CMIF}_k(j\omega) =\sigma_{k}(j\omega)^2 \, .
\end{equation}
Eq. \eqref{eqn: CMIF} is then useful for additional cross-validation among the models. 

Lastly, the SVD extraction is used to quantify the relative distance of the models at each frequency using 
\small{\begin{equation}
\label{eqn: errore}
    \%||\Delta \boldsymbol{T}(j\omega)||_{\infty} =\frac{\sup_{\omega\in\mathbb{R}}\sigma_{\max} (\boldsymbol{T}_\text{BB}(j\omega)-\boldsymbol{T}_\text{GB}(j\omega,k^*))}{ \sup_{\omega\in\mathbb{R}}\sigma_{\max} (\boldsymbol{T}_\text{BB}(j\omega))}\times 100 \, ,
\end{equation}
}
where the closed loop MIMO transfer function of the BB is $T_\text{BB}(j\omega)$, and for GB is $T_\text{GB}(j\omega)$. \eqref{eqn: errore} is similar to \eqref{eqn: Hinfty} but accounts for differences between the closed-loop transfer functions.

% Lastly, the SVD extraction is used to quantify the relative distance of the models for each frequency, as shown in \label{eqn: errore} by using \eqref{eqn: Hinfty}, but on the difference between the closed loop transfer functions as shown in \eqref{eqn: errore}. Where closed loop MIMO transfer fucntion of the BB $T_{\text{BB}}(j\omega)$ and GB $T_{\text{BB}}(j\omega)$.

\subsection{Type-$f$ and Type-$\zeta$ mismatch}
Quantifying the divergence with one of the two SVD-based methods allows grasping different aspects of how two models differ. First, CMIF can identify the dominant mode frequency and the damping effect, but it cannot furnish a trustworthy value in relative terms due to the reverse triangular inequality defined by:
\begin{equation}
\label{eqn: traing}
\left| \|G_\text{A}\|_{\infty} - \|G_\text{B}\|_{\infty} \right| \le \|G_\text{A} - G_\text{B}\|_{\infty} \, ,
\end{equation}
where $G_A$ and $G_B$ are the converter and Test Grid transfer function, respectively. Given the fact that $H_{\infty}$ represents the superior values of singular values at each frequency, the straight comparison of the CMIF would result in an underestimation of the potential difference between the models. Second, the evaluation of the relative FD error in \eqref{eqn: errore} is hardly sensitive to frequency responses, even mildly shifted. This creates large FD errors that may be due to minor divergence of corner frequencies. 

\begin{figure}[t!]
    \centering
    \hspace{-5mm}
    \includegraphics[width=1\linewidth]{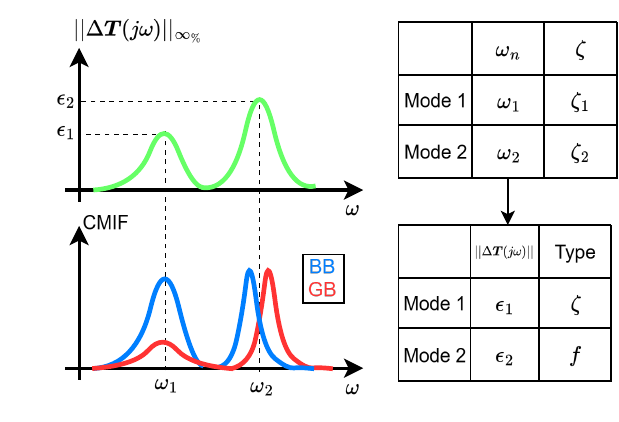}
    \vspace{-3mm}
    \caption{Quantification and qualification of the gray-box model's error.}
    \label{fig: quanti}
\end{figure}

Therefore, a combination of both CMIF and \eqref{eqn: errore} could be used to quantify the divergence as shown in \figref{fig: quanti}. More specifically, they could be combined to furnish a tolerance interpretation on the modal analysis performed with GB models. In this regard, the research identifies two principal types of divergence: frequency mismatch (Type-$f$) and damping mismatch (Type-$\zeta$). The mismatch could also be because of the mix of types.

CMIF highlights dominant energetic modal contributions, while SSA estimates poles, including weakly observable ones. Combining CMIF and SSA provides insight into the FD representation of the entire system. The SSA estimates the modes' damping and frequency, while CMIF also includes the most energetic directions by including the residues. Interpreting both CMIF peaks and modes helps distinguish between theoretically present modes and those that affect the observed response, especially in MIMO systems, where control can attenuate or amplify modal contributions based on residues.

\section{Results}
\label{results}
%% Intro to the results section linking methodology used and cases
This section demonstrates how the gray-boxing framework performs when applied to increasingly uncertain cases, ranging from 100\% open to the 0\% open case represented by the detailed EMT BB model (with generic non-project specific parameters), using a white-box model to illustrate the EMT scheme shown in 
\figref{fig: WB-structure-SSM}. The cases studied will be:

\begin{itemize}
    \item Case 1: Parametric recovery
    \item Case 2: Structure mismatch
    \item Case 3: Detailed Black-Box
\end{itemize}

The test cases are designed to gradually test the framework and identify and evaluate its effectiveness and inefficiencies. Therefore, for each case, the time response equivalence between the GB and BB, the extracted gains that make the GB similar to the BB, and the optimizer process information will be shown. Lastly, the distance in FD is also quantified for each case. 

\paragraph{Test cases configuration}

Referring to the setup configuration procedure shown in \figref{fig: simulation_setup} and variables from the WB in \figref{fig: WB-structure-SSM}, the setup conditions for Test Grid, I/O signals, and elements in the WB are expressed in \tabref{tab:test_signals_config}. Additionally, every passive parameter of the BBs cases is placed in the WB models in the same way. 

\begin{table}[t!]
\centering
\caption{Simulation configuration for all cases.}
\label{tab:test_signals_config}
\begin{tabular}{m{3cm} c c c}
\toprule
Item & Case 1& Case 2.1/2.2 & Case 3\\
\midrule
Time step, $\Delta t$ & $\SI{10}{\micro\second}$ & $\SI{10}{\micro\second}$ & $\SI{10}{\micro\second}$ \\
\addlinespace
$u_1(t)$: signal & $P_{\text{ref}}$ [pu] & $P_{\text{ref}}$ [pu] & $\theta_{\text{grid}}$ [deg]\\
$u_1(t)$: step & $0\rightarrow0.5$ & $0\rightarrow0.5$ & $0\rightarrow27$ \\
$u_1(t)$: instant [s] & $2$ & $2$ & $4$ \\
$u_2(t)$: signal & $ V_{\text{ref}}$ [pu] & $ V_{\text{ref}}$ [pu] & $ V_{\text{grid}}$ [pu]\\
$u_2(t)$: step & $1\rightarrow0.95$ & $1\rightarrow0.95$ & $1\rightarrow0.95$ \\
$u_2(t)$: instant [s] & $4$ & $4$ & $6$ \\
\addlinespace
$y_{\text{out}_1}$ & $v_{\text{abc}}$ & $v_{\text{abc}}$ & $P_{\text{AC}}$ \\
$y_{\text{out}_2}$ & $i_{\text{abc}}$ & $i_{\text{abc}}$ & $V_{\text{PoC}}$ \\
\addlinespace
Machine dynamics & no & no/yes & yes \\
Lag‑lead filter & no & yes  & yes \\
\addlinespace
Short-circuit ratio (SCR) & $3$ & $3$ & $3$ \\
$X/R$ & $7$ & $7$ & $7$ \\
Rated rms line-to-line voltage, $V_\text{b}$ [kV]& $220$ & $220$ & $0.69$ \\
Rated apparent power, $S_b$ [MVA] & $120$ & $120$ & $10$\\
\bottomrule
\end{tabular}
\end{table}

% Case 1: Same model
\subsection{Case 1: Same model structure}
\label{sub: parametric recovery}
\begin{figure*}[t!]
    \centering
    \includegraphics[width=0.9\linewidth]{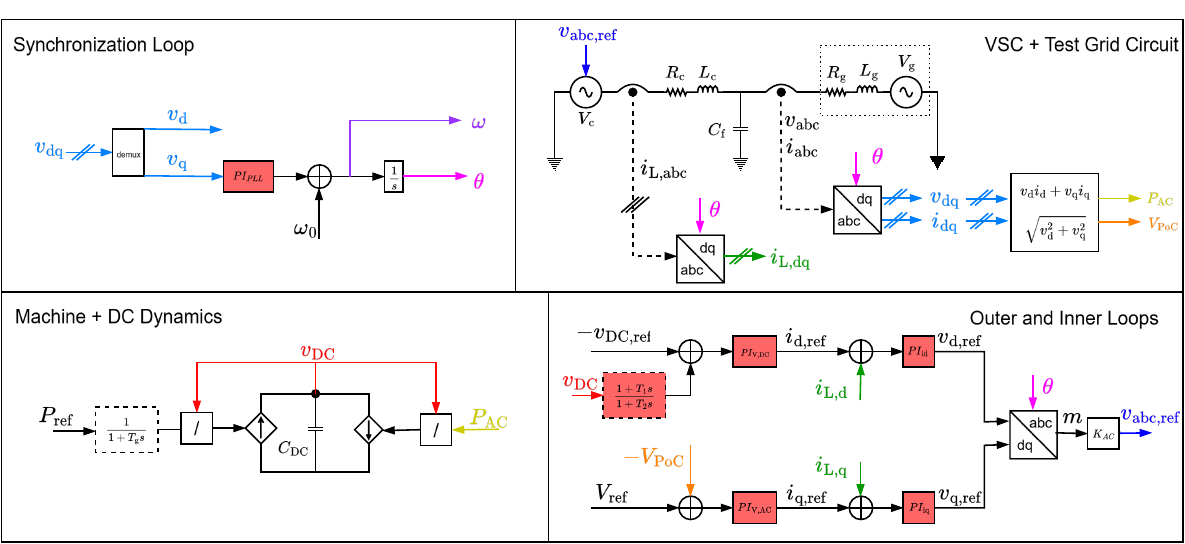}
    \caption{White-box ansatz with degrees of freedom highlighted (red boxes with continuous lines).}
    \label{fig: WB-structure-SSM}
\end{figure*}

The first test assumes full knowledge of the BB structure and parameters. For the GB structure (a simplified version of \figref{fig: WB-structure-SSM}, without the elements highlighted with dotted lines in the figure, i.e., the lag-lead filter and the machine dynamics), it is assumed to be fully known and equal to the BB’s, except for the (gain) parameters, which are identified through the gray-boxing method and compared with their nominal values. Therefore, in this case, the framework's ability to recover the same gains when the structure matches is tested, and accuracy is shown in \tabref{tab:gains_comparison}.

\begin{table}[t!]
\centering
\caption{Optimization results for Case 1.}
\label{tab:gains_comparison}
\begin{tabular}{lccc}
\toprule
Gain [p.u.] & Initial & Original & Case 1 \\
\midrule
$k_\text{p,V}$  & 1   & 0       & 0.002543 \\
$k_\text{i,V}$  & 600 & 555.59  & 555.49 \\
$k_\text{p,P}$  & 1   & 5.0943  & 5.0127 \\
$k_\text{i,P}$  & 600 & 539.43  & 539.51 \\
$k_\text{p,Id}$ & 1   & 2.7167  & 2.7081 \\
$k_\text{i,Id}$ & 600 & 3266.8  & 3229.2 \\
$k_\text{p,Iq}$ & 1   & 2.7167  & 2.7085 \\
$k_\text{i,Iq}$ & 600 & 3266.8  & 3290.8 \\
$k_\text{p,PLL}$ & 2  & 5.7003  & 5.7643 \\
$k_\text{i,PLL}$ & 2  & 3.1342  & 4.0481 \\
\midrule
\multicolumn{4}{l}{Obj = $5.2925\times10^{-5}$, Elapsed time = 3469.09 s.} \\
\bottomrule
\end{tabular}
\end{table}

\begin{figure*}[t!]
    \centering
    \begin{subfigure}[t!]{0.24\textwidth}
        \centering
        \includegraphics[width=\linewidth]{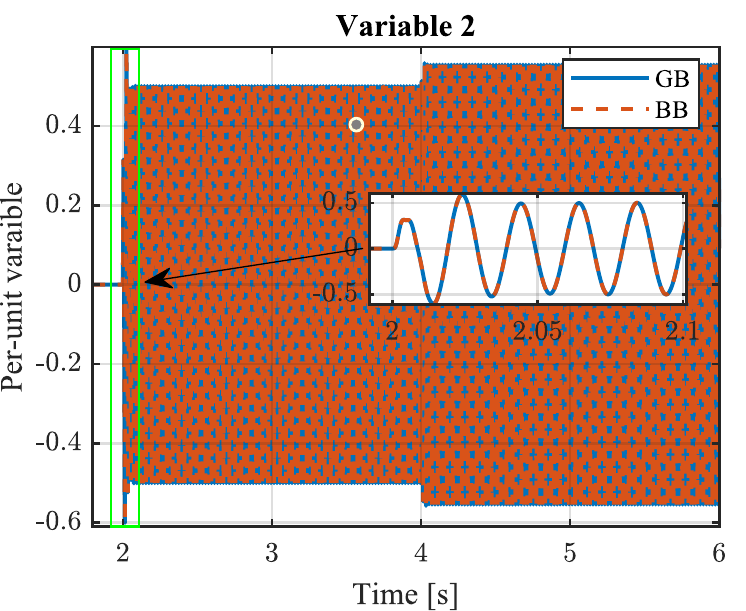}
        \caption{Time-domain}
        \label{fig:plot1_case1}
    \end{subfigure}
    \begin{subfigure}[t!]{0.24\textwidth}
        \centering
        \includegraphics[width=\linewidth]{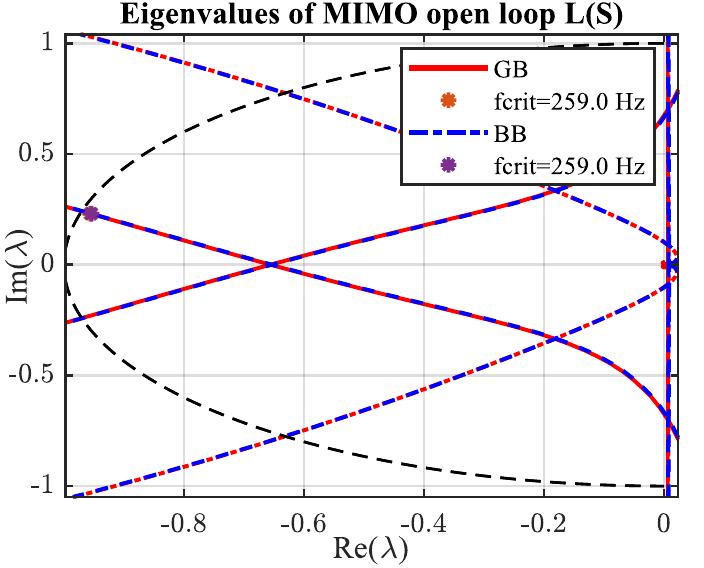}
        \caption{Nyquist}
        \label{fig:plot2_case1}
    \end{subfigure}
    \begin{subfigure}[t!]{0.24\textwidth}
        \centering
        \includegraphics[width=\linewidth]{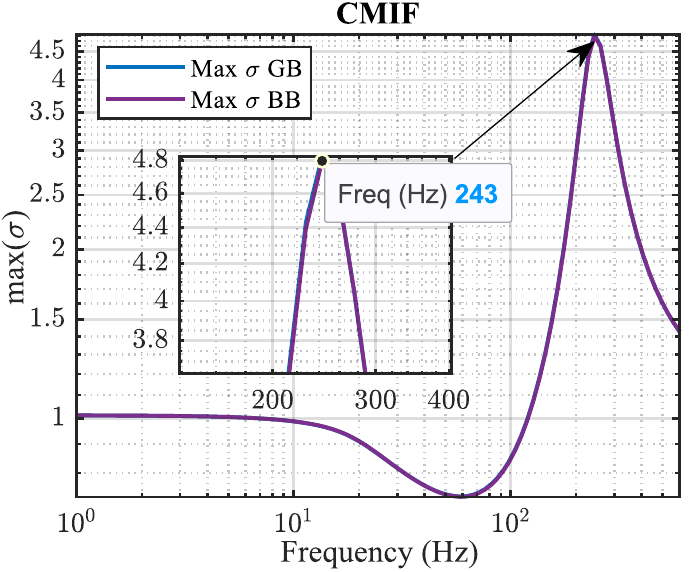}
        \caption{CMIF}
        \label{fig:plot3_case1}
    \end{subfigure}
    \begin{subfigure}[t!]{0.24\textwidth}
        \centering
        \includegraphics[width=\linewidth]{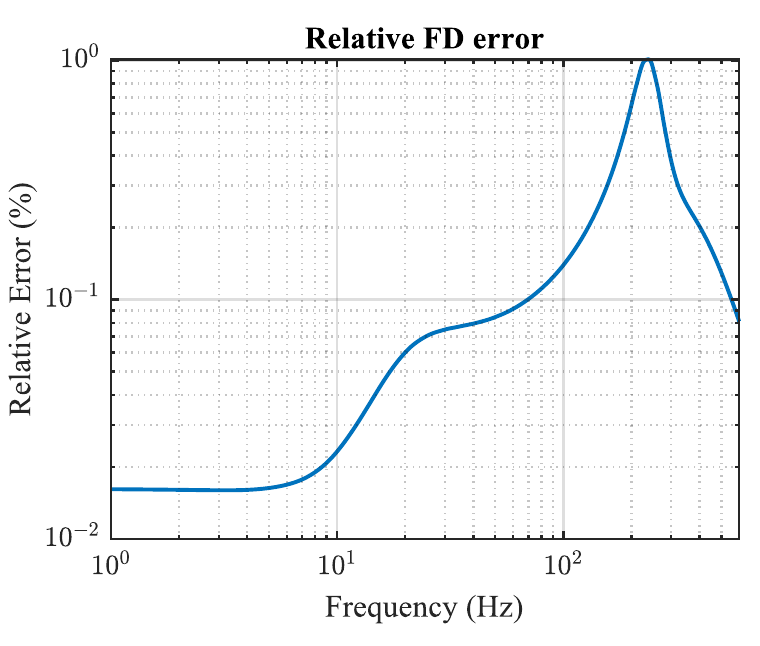}
        \caption{FD Error}
        \label{fig:plot4_case1}
    \end{subfigure}
    \caption{Case 1 results.}
    \label{fig:caso1}
\end{figure*}

A notable aspect is the method's ability to capture exact gains and time response in \figref{fig:plot1_case1} when the control structure is equivalent, as well as its relatively short time requirement. Since the gray-boxing framework caught the exact initial gains, it is worth only highlighting how the (visual) Nyquist criterion displays the frequency of the most undamped mode, ${\omega_{\text{d},k} | \min{\zeta_k}}$, inaccurately as shown in \figref{fig:plot2_case1}. Therefore, this results in a coarse method for such a task. Hence, it would be better either to use the CMIF, shown in \figref{fig:plot3_case1}, about the ${\omega_{\text{d},k} | \min{\zeta_k}}$. Lastly, for the sake of illustrating how the FD-distance metric \eqref{eqn: errore} performs in this case, it is shown in \figref{fig:plot4_case1}. It is clear that the difference in FD is negligible since it is practically less than $1\%$ for the whole analyzed frequency range; however, the justification for the minor mismatch can arise from the non-perfect matching of the extracted gains in \tabref{tab:gains_comparison},

% FIGURA CASO 1 - 4 plot in riga

The eigenvalue analysis comparison is shown in \tabref{tab:less_damped_modes}, where the GB SSM-Twin eigenvlaues are displayed against the ones extracted through vector fitting the frequency response of the BB. Additionally, it is done on the and the participation factor (PF) on the ${\omega_{\text{d},k} | \min{\zeta_k}}$ thanks to the GB SSM-Twin is shown in \tabref{tab:less_damped_modes PF}, also highlighting the matching with the peak frequency in CMIF, approximately $\SI{243}{Hz}$ (\figref{fig:plot3_case1}).

\begin{table}[t!]
\centering
\caption{Critical modes for Case 1.}
\label{tab:less_damped_modes}
\begin{tabular}{cccc}
\hline
Case & Mode ID & Damping [\%] & Frequency [Hz] \\
\hline
\multirow{3}{*}{Analytical SSM} 
& 7 & 16.491 & 241.77 \\
& 5 & 55.847 & 310.95 \\
& 9 & 59.552 & 16.44 \\
\hline
\multirow{3}{*}{Fitted SSM} 
& 18 & 16.72 & 238.58 \\
& 12 & 55.8 & 307.79 \\
& 7 & 59.212 & 16.46 \\
\hline
\end{tabular}
\end{table}

\begin{table}[t!]
\centering
\caption{Participation factor (PF) of most undamped modes, ${\omega_{\text{d},k} | \min{\zeta_k}}$, for the gray-box small-signal model twin in Case 1.}
\label{tab:less_damped_modes PF}
\begin{tabular}{ccc}
\hline
State Name & State Symbol  & PF [\%] \\
\hline
Grid $\mathrm{q}$-axis current, system frame & $i_\text{g,q}^\text{s}$ & 63.57 \\
Grid $\mathrm{d}$-axis current, converter frame &$i_\text{g,d}^\text{c}$ & 33.70 \\
VSC $\mathrm{q}$-axis current, system frame &$i_\text{L,q}^\text{s}$ & 1.01 \\
\hline
\end{tabular}
\end{table}

% Case 2: Slightly different structure
\subsection{Case 2: Structure mismatch} 
\label{sub: structure mismatch}
The second case tests the gray-boxing framework's ability to accommodate differences in control structure. More specifically, in this case, two degrees of freedom ($T_1$, $T_2$) were added through a lag-lead filter in the GB, shown in \figref{fig: WB-structure-SSM}, which are expected to attenuate the structural differences. The filter is placed on the d-axis control, particularly on the DC capacitor's voltage measurement, since sensitivity analysis showed that d-axis control parameters influence the frequency response the most \cite{Mugambi2025Impedance-basedGenerator}. Whereas the structural differences in the BB version, in addition to the different gains, would have:

\begin{itemize}
    \item A first-order filter \eqref{eqn: machine} to mimic the additional machine side dynamics
    \begin{equation}
    \label{eqn: machine}
        G_\text{m}(s) = \frac{1}{0.1s+1}
    \end{equation}
    \item A different Normalized-PLL structure where $v_\text{q}'=v_\text{q}/v_\text{d}$
    \item Feed forward and current decoupling on the inner control loops
\end{itemize}

Notice that some structural differences are control-based, while the machine dynamics represent a machine-based difference. Therefore, the extraction was performed both with and without machine dynamics to assess the differences in identification capability, resulting in Case 2.1 (GB) and Case 2.2 (GB+), respectively.
The equivalent gains extracted, compared to the known BB ones, are listed in \tabref{tab:gains_comparison_2}.

\begin{table}[t!]
\centering
\caption{Optimization results for Case 2.}
\label{tab:gains_comparison_2}
\begin{tabular}{lcccc}
\toprule
Gain [p.u.] & Initial & Original & Case 2 & Case 2.1 \\
\midrule
$k_\text{p,V}$  & 1   & 0       & 0       & 0.03645 \\
$k_\text{i,V}$  & 600 & 555.59  & 596.24  & 596.72 \\
$k_\text{p,P}$  & 1   & 5.0943  & 0.47706 & 5.5639 \\
$k_\text{i,P}$  & 600 & 539.43  & 0.00778 & 599.93 \\
$k_\text{p,Id}$ & 1   & 2.7167  & 10      & 11.435 \\
$k_\text{i,Id}$ & 600 & 3266.8  & 992.57  & 1000 \\
$k_\text{p,Iq}$ & 1   & 2.7167  & 1.6023  & 1.4245 \\
$k_\text{i,Iq}$ & 600 & 3266.8  & 904.47  & 1000 \\
$k_\text{p,PLL}$ & 2  & 5.7003  & 12.177  & 5 \\
$k_\text{i,PLL}$ & 2  & 3.1342  & 3       & 3.6439 \\
$T_1$           & 0.1 & --      & 0.035692 & 0.88852 \\
$T_2$           & 0.1 & --      & 0.052732 & 1 \\
\midrule
\multicolumn{5}{@{}l@{}}{\textbf{Obj 1 = $2.6532$, Elapsed time 1 = \SI{4228.57}{s}.}} \\
\multicolumn{5}{@{}l@{}}{\textbf{Obj 2 = $0.01355$, Elapsed time 2 = \SI{3423.51}{s}.}} \\
\bottomrule
\end{tabular}
\end{table}

The first notable aspect of the synthesized gains is how the nested control structure, in addition to the lag-lead filter, adapts to absorb the differences in the control structure. This is even more significant for the GB without machine dynamics in Case 2.2, where the gains, principally the outer loop's d-branch quantities, are considerably smaller. More specifically, the outer loop's ability to maintain constant DC dynamics is drastically reduced to a non-aggressive DC-voltage droop-based control. While the GB with the machine dynamics in Case 2.1 captures the original control gains' order of magnitude. 

\begin{figure*}[t!]
    \centering
    \begin{subfigure}[t!]{0.24\textwidth}
        \centering
        \includegraphics[width=\linewidth]{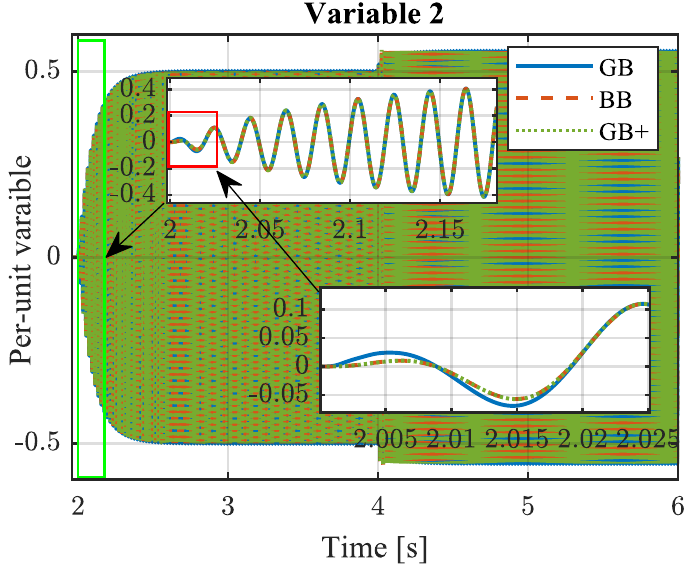}
        \caption{ot Time-domain}
        \label{fig:plot1_case2}
    \end{subfigure}
    \hfill
    \begin{subfigure}[t!]{0.24\textwidth}
        \centering
        \includegraphics[width=\linewidth]{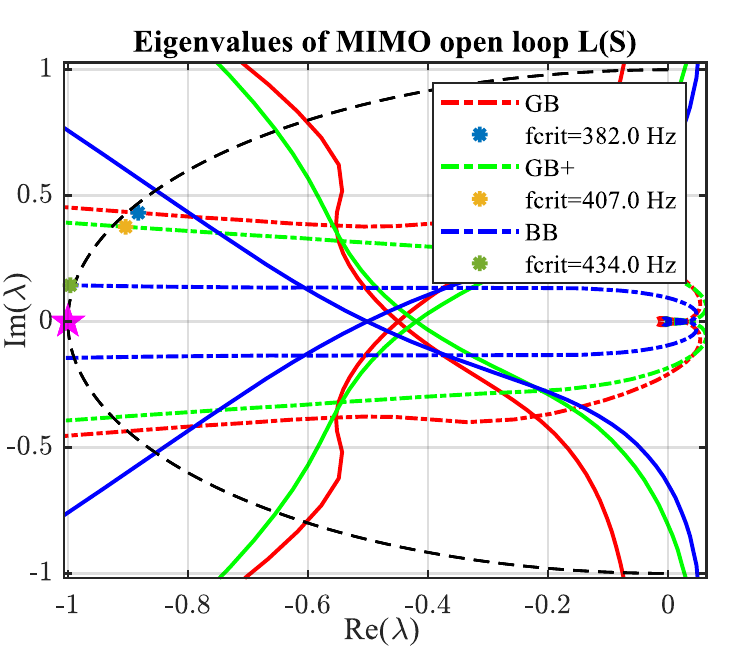}
        \caption{Nyquist}
        \label{fig:plot2_case2}
    \end{subfigure}
    \hfill
    \begin{subfigure}[t!]{0.24\textwidth}
        \centering
        \includegraphics[width=\linewidth]{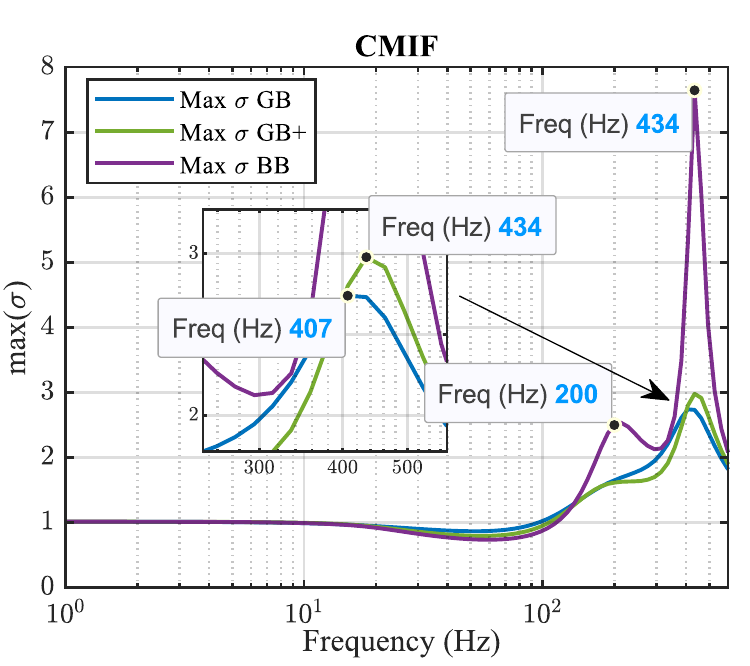}
        \caption{CMIF}
        \label{fig:plot3_case2}
    \end{subfigure}
    \hfill
    \begin{subfigure}[t!]{0.24\textwidth}
        \centering
        \includegraphics[width=\linewidth]{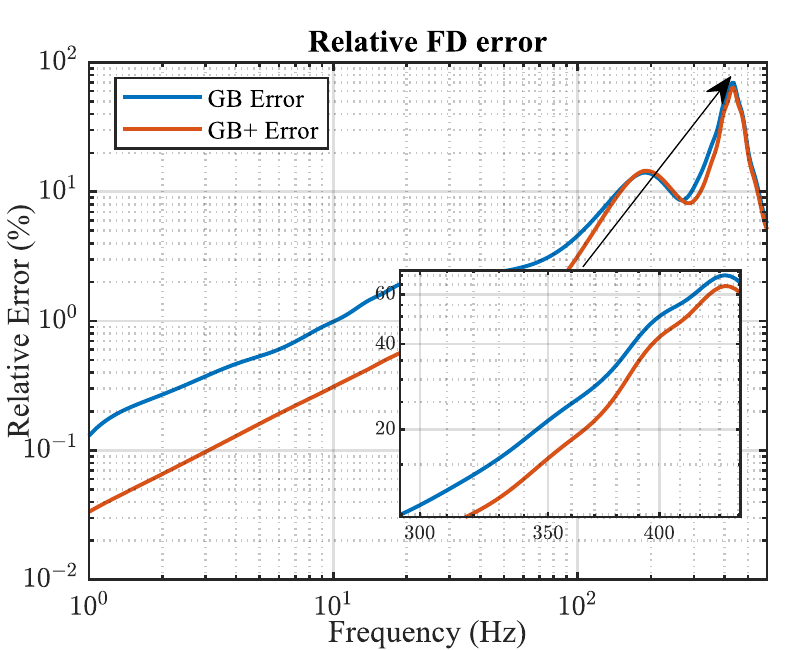}
        \caption{FD Error}
        \label{fig:plot4_case2}
    \end{subfigure}
    \caption{Case 2 results.}
    \label{fig:caso2}
\end{figure*}

As concerns the objective function, it is clear visually in \figref{fig:plot1_case2} and numerically in \tabref{tab:gains_comparison_2} that the extraction method captures well the BB behavior, with particular precision and speed if the GB also has the machine dynamics.

In FD, the main difference is that the peak of the frequency response shifts to lower frequencies in the GB case to simulate the machine slowness \eqref{eqn: machine}, simply by adjusting the controls in \figref{fig: WB-structure-SSM}. This is achieved through non-symmetrical control, where the inner-loop d-gains can differ from the q-gains. This helps to compensate for the differences, functioning as four DoFs instead of two.

As concerns the ${\omega_{\text{d},k} | \min{\zeta_k}}$, it is possible to notice graphically in the Nyquist plot in \figref{fig:plot2_case2} and in the CMIF in \figref{fig:plot3_case2}, that it was not precisely captured from the extracted concept model. By performing the SSA on the BB and the GB SSM Twin model, the difference in damping and natural frequencies of the most significant modes are shown in \tabref{tab:case2 SSA}. More specifically, in the GB without the machine, it was not possible to precisely control the frequency or the damping of ${\omega_{\text{d},k} | \min{\zeta_k}}$. Additionally, the GB+ model with the machine captures the frequency and damping exactly at the minimum, but in absolute terms, it quantifies them twice as much as the BB case.

It is worth noting that the accuracy of the modes found through modal analysis can be quantified by combining the CMIF and relative FD error plots in \figref{fig:plot4_case2}. By doing so, it is possible to define that the ${\omega_{\text{d},k} | \min{\zeta_k}}$ identified with GB and GB+ are in a frequency range where they have both $\approx60\%$ divergence compared to the BB. In addition, the GB case exhibits both Type-$f$ and Type-$\zeta$ divergences, whereas GB+ shows only Type-$\zeta$.

\begin{table}[t!]
\centering
\caption{Critical modes for Case 2.}
\label{tab:case2 SSA}
\begin{tabular}{cccc}
\hline
Case & Mode ID & Damping [\%] & Frequency [Hz] \\
\hline
\multirow{3}{*}{Analyt. SSM Case 2.1} 
& 4 & 19 & 414 \\
& 6 & 60 & 130 \\

\hline
\multirow{3}{*}{Analyt. SSM Case 2.2} 
& 4 & 12 & 433 \\
& 6 & 50 & 143 \\

\hline
\multirow{3}{*}{Fitted SSM} 
& 2 & 6.80 & 434 \\
& 1 & 27 & 188 \\

\hline
\end{tabular}
\end{table}

% Case 3: Complex commercial model

\subsection{Case 3: Detailed Black-Box}
\label{sub: Commercial Black-Box}
The last case tests the limits of the framework with one of the most challenging scenarios: a detailed switching EMT model with full-order converter control setup is synthesized into GB. In this scenario, the GB setup is evaluated on an actual model, and the difference between the EMT's real BB model and the extracted one is measured. Since the detailed BB model is a full-order model, it is well-suited for disturbance rejection studies. This type of model includes real-life practical limitations and is characterized by large ramp-rate limiters on certain reference signals, so they cannot be universally used as $	\textbf{u}_\text{ref}$ because the rate limiter would mask the internal dynamics, biasing the final results. 
To still excite the active power and voltage controls, the perturbation is alternatively applied on the Test Grid side $\theta_\text{grid}(t) = \ang{27} H(t-\SI{4}{s})$ ,
$V_\text{grid}(t) = \left[1 - 0.05H(t-\SI{6}{s}) \right]\si{pu}$ as shown in \tabref{tab:test_signals_config}. 

The extracted control gains from the detailed BB model are shown in \tabref{tab:gains_comparison_3_rescaled}. In this case, the selection of the initial gains was designed so that the optimizer would follow a Ziegler-Nichols PID tuning technique \cite{Dormido2015PIDControl}, starting with low proportional and integral gains. In this case, the result is a local minimum, with the objective/cost function stopping at the convergence value of 45.89 due to the accumulation of distance from the switching effect of the OEM's WT outputs. 

\begin{table}[t!]
\centering
\caption{Optimization results for Case 3}
\label{tab:gains_comparison_3_rescaled}
\begin{tabular}{lccc}
\toprule
Gain [p.u.] & Initial & Case 3 \\
\midrule
$k_\text{p,V}$   & 0.5   & 0.01102  \\
$k_\text{i,V}$   & 1     & 7.463    \\
$k_\text{p,P}$   & 0.5   & 0.4330   \\
$k_\text{i,P}$   & 1     & 1.921    \\
$k_\text{p,Id}$  & 0.5   & 1.905    \\
$k_\text{i,Id}$  & 100   & 61.245   \\
$k_\text{p,Iq}$  & 0.5   & 0.9919   \\
$k_\text{i,Iq}$  & 100   & 44.75    \\
$k_\text{p,PLL}$ & 6     & 14.27    \\
$k_\text{i,PLL}$ & 4     & 2.950    \\
$T_1$            & 1E-5  & 0.00098033 \\
$T_2$            & 0.001 & 0.0014462  \\
\midrule
\multicolumn{3}{@{}l@{}}{Obj = $45.89$, Elapsed time = 1222.94 s.} \\
\bottomrule
\end{tabular}
\end{table}

\begin{figure*}[t!]
    \centering
    \begin{subfigure}[t!]{0.24\textwidth}
        \centering
        \includegraphics[width=\linewidth]{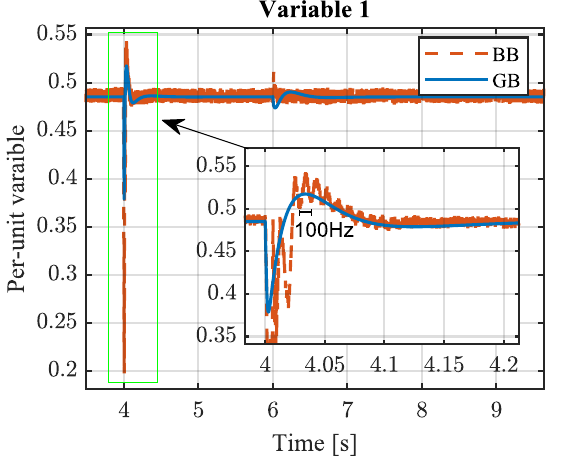}
        \caption{Time-domain}
        \label{fig:plot1_case3}
    \end{subfigure}
    \begin{subfigure}[t!]{0.24\textwidth}
        \centering
        \includegraphics[width=\linewidth]{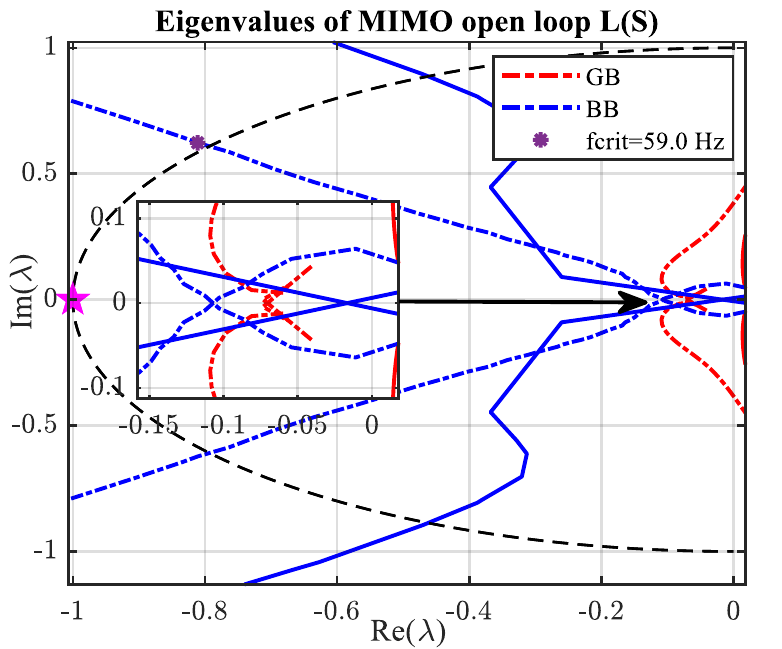}
        \caption{Nyquist}
        \label{fig:plot2_case3}
    \end{subfigure}
    \begin{subfigure}[t!]{0.24\textwidth}
        \centering
        \includegraphics[width=\linewidth]{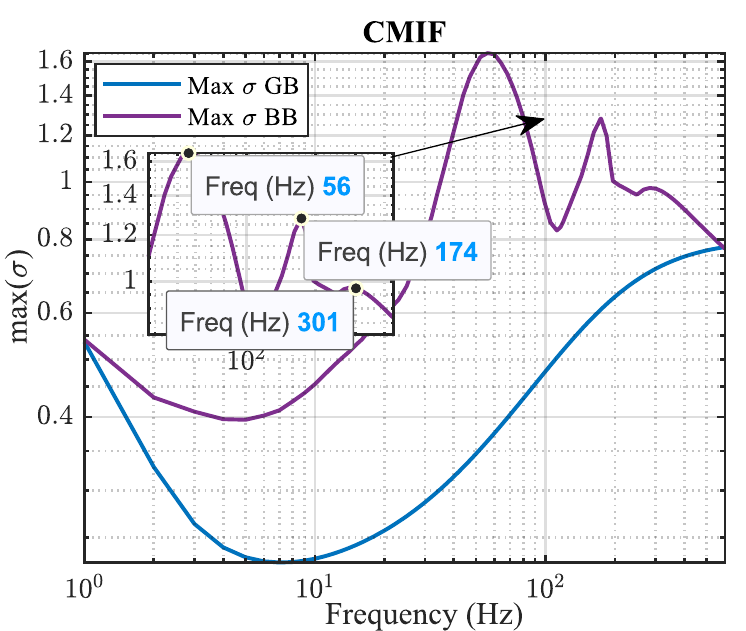}
        \caption{CMIF}
        \label{fig:plot3_case3}
    \end{subfigure}
    \begin{subfigure}[t!]{0.24\textwidth}
        \centering
        \includegraphics[width=\linewidth]{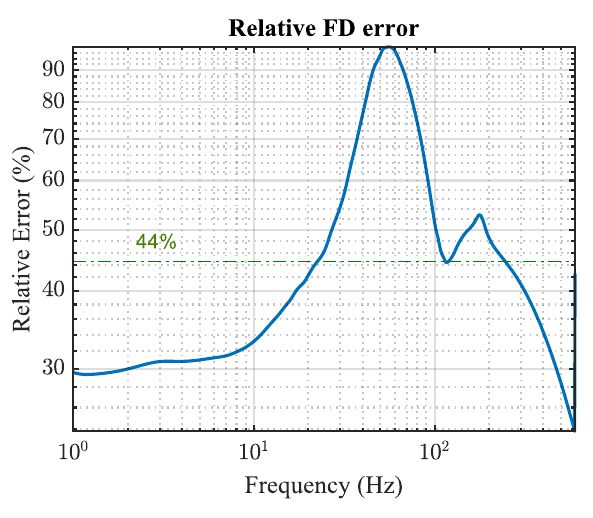}
        \caption{FD Error}
        \label{fig:plot4_case3}
    \end{subfigure}
    \caption{Case 3}
    \label{fig:caso3}
\end{figure*}

In \figref{fig:caso3} are shown the results of the extraction in time and FD. In particular, the time-domain active power outputs of the commercial model are compared with GB's in \figref{fig:plot1_case3}. The GB replicates the BB in terms of time constant and overshoot. However, it cannot capture additional high-frequency oscillations. On top, there are $\SI{100}{Hz}$ oscillations that are not replicated in the GB model. These represent negative-sequence currents during the phase-shift transient, which the detailed model can control and manage through additional negative-sequence control branches, a feature that is not present in the GB model. 

The GB model, as shown in \figref{fig: WB-structure-SSM} and used for Cases 1 and 2, once properly tuned, behaves like an average of the detailed model; this is also evident from the modes, where the principal modes shown in \tabref{tab:case3 SSA} are not captured. This is also visible in \figref{fig:plot3_case2} and \figref{fig:plot3_case3} where it is possible to see that the frequencies with the most energetic input-output are not corresponding to the most undamped modes, indicating that, despite being the most undamped modes, they are also associated with a low residual factor, thus having a low representation in the measured output. 

The distance from the models is shown in \figref{fig:plot4_case3} with an average divergence of the GB's model from the OEM of 44\%, additionally indicating that in the $\left[1,600\right]\si{Hz}$ range for every mode found there is a Type-$f,\zeta$ divergence with the corresponding relative error. 

\begin{table}[t!]
\centering
\caption{Most significant modes for Case 3}
\label{tab:case3 SSA}
\begin{tabular}{cccc}
\hline
Case & Mode & Damping [\%] & Frequency [Hz] \\
\hline
\multirow{3}{*}{Analytical SSM} 
& 3 & 9 & 3643 \\
& 1 & 16 & 3572 \\

\hline
\multirow{3}{*}{Fitted SSM} 
& 8 & 5 & 106 \\
& 6 & 12 & 188 \\

\hline
\end{tabular}
\end{table}

This represents the potential limit case where the method proposed in the paper is benchmarked against the detailed EMT model. From the results, it is possible to conclude that the detailed model in question has a potentially significantly different structure from the one assumed for the GB (feed-forward loops, delays, switching, decoupling, parallel control structures for negative sequence, etc.). 

Therefore, using a simplified surrogate model for stability analysis may yield biased results. The proposed method can, on one hand, provide a way to quantify and qualify the reliability of the analysis; on the other hand, it can help to adapt the GB structure (e.g., addition of negative sequence controls) to better replicate the BBs and reduce their distance in FD under the threshold deemed acceptable within the required frequency range.

\section{Discussion}
\label{discussion}
The proposed EMT gray-boxing framework addresses whether it is possible to extract physically meaningful parameters from black-box IP-protected models and examines their implications and limitations. 

\subsection{Framework Performance and Key Insights}
From the results of the gradually more complex cases, starting from the easy parameter recovery, arriving at the OEM's model extraction, which represents the edge case, it was possible to derive three principal key insights:

\begin{itemize}
    \item Structural equivalence leads to near-perfect recovery (Sec. \ref{sub: parametric recovery}): 
    
    Once the initial structure is known, it is possible to recover the exact initial VSC gains with an FD divergence of less than 1\%.
    \item Cascade control with DoF compensates for moderate mismatch (Sec. \ref{sub: structure mismatch}): 
    
    A nested control structure with additional DoF, such as a filter, can absorb structural differences. When physical elements are missing in the GB model compared to the BB version, the controls will compensate by assuming physically unrealistic values.
    \item FD divergence quantification ensures SSA reliability (Sec. \ref{sub: Commercial Black-Box}): 
    
    The FD error and Type-$f/\zeta$ metrics provide tolerance bounds and a measure of divergence, linking the SSA results to a fidelity-check quantification criterion. This is relevant when the internal GB structure does not match the BB structure; thus, an intrinsic, unavoidable difference exists and must be accounted for.
\end{itemize}

\subsection{Situating the results in the existing literature}

Extracting stability information from frequency responses is limited to eigenvalues and may yield spurious poles, lacking insight into internal structures. Methods that adjust the FD model for faster analysis can optimize parameters but may introduce mismatches in EMT models. Nonlinear FD elements have been explored, but time-domain step responses capture a wider range of operational points and align better with open models. However, fitting a black-box model to a different structure may not retain all aspects of the original model. While structural adjustments can yield physically unrealistic gains, known structures can be accurately reconstructed using terminal measurements. The research aims to quantify the distance between FD models to analyze bias in study results.

\subsection{Practical Implications}
The primary output of the research is the EMT GB framework, which systematically extracts controller gains from various models and structures applicable to open standard models. These gains will emulate the initial time-response through clearly defined phases: setup, extraction, and divergence quantification.

Prior knowledge of the system structure greatly improves the identification process, ensuring accurate parameter recovery when the initial model is correct. However, as structural uncertainty grows, the uniqueness of the solution may decrease. In such cases, the extracted parameters can still produce a good fit in the time-domain but may show noticeable differences in the FD. This is important: it suggests that revealing the exact initial model isn't strictly necessary, as long as the identified model meets predetermined FD tolerance criteria, which is effectively verified through our proposed quantification framework.

 The framework’s divergence quantification enables the use of standard models, yielding results similar to the original and providing insights into the surrogate model’s accuracy, helping assess the fidelity of the outcomes. It clarifies whether discrepancies arise from damping, natural frequency mismatches, or both.

Overall, the research aims to enable the safe integration of IP-protected IBRs into the power system by facilitating the synthesis of realistic IBR models for studies. This supports open SSA execution and uncertainty quantification, providing insights into the factors that contribute to specific events, even at the physical control level.

The straightforward time-domain data inputs for the gray-boxing framework make it suitable for users such as TSOs, emphasizing ease of use.

\subsection{Limitations and future work}

The limitations encountered are mainly related to the optimization process, the selection of the optimizer, and the initial gain guess. The extraction accuracy and speed are significantly sensitive to these aspects. The research's scope relates to the general gray-boxing framework definition, but it is worth noting that there is room to explore how different optimization methods perform. 

The research acknowledges that if meaningful physical values are desired for the gray-boxing structure, some differences, principally visible in the FD, must be accepted. Generally, surrogate models are suitable for limited frequency ranges; therefore, if the extracted model results show unacceptable quantitative divergence, the research suggests simply performing several iterations by varying the GB structure to fit the model within the desired frequency range and tolerance. 

In this regard, future work may consider adding a hybrid EMT and FD to ensure that your GB has both physically meaningful values and the desired distance at the target frequency range. 

\section{Conclusion}
\label{conclusion}
The paper introduces a systematic gray-box framework for identifying voltage source converter (VSC) models from terminal time-series data, relying only on black-box measurements and a physically based white-box ansatz. It explains that similarity in the time-domain does not ensure equivalence in the frequency-domain (FD), and therefore adds a complementary FD metric to measure model divergence and differentiate between frequency mismatch and damping mismatch. The findings show perfect recovery when the internal structure is known, acceptable compensation with moderate structural mismatch, and highlight the importance of interpreting surrogate-model-based SSA alongside a fidelity measure in more uncertain situations. The framework is especially useful for IP-protected converter models, where it can help create physically meaningful surrogate models and provide a confidence level for their application in stability analysis.
\section{Acknowledgements}
This work is supported by the European Union as part of ADOreD project funded by the Horizon Europe MSCA programme (\href{https://www.msca-adored.eu/}{HORIZON-MSCA-2021-DN, Grant agreement 101073554})

\section{Legal Disclaimer}
Figures and values presented in this paper should not be used to judge the performance of Siemens Gamesa Renewable Energy technology as they are solely presented for demonstration purpose. Any opinions or analysis contained in this paper are the opinions of the authors and not necessarily the same as those of Siemens Gamesa Renewable Energy.

%Bibliography
\bibliographystyle{unsrt}  
\bibliography{references}

\end{document}